# Analysis of diagonal *G* and subspace *W* approximations within fully self-consistent *GW* calculations for bulk semiconducting systems

Yashpal Singh† and Lin-Wang Wang‡

*Material Science Division, Lawrence Berkeley National Laboratory, Berkeley, California, 94720, USA*

Fully self-consistent *GW* (*sc-GW*) methods are now available to evaluate quasiparticle and spectral properties of various molecular and bulk systems. However, such techniques based on the full matrix of *G* and *W* are computationally demanding. The routinely used single-shot *GW*-approximation ($G_0W_0$) has an undesirable dependency on the choice of initial exchange-correlation functional. In the literature, many so-called self-consistent *GW* methods are based on diagonal approximation of *G* and low-ranking approximation of W. It is thus worth to check how good such approximations are in comparison with the full matrix method. In this work, we consider AlAs, AlP, GaP, and ZnS as the prototype systems to perform *sc-GW* calculations by expressing the full *G* matrix using a plane-wave basis set. We compared our *sc-GW* results with the diagonal *G* and subspace *W* approximated *sc-GW results* (*sc-GW-diagG* and *sc-GW-subW* methods). In the *sc-GW-diagG* method, interacting *G* is expanded in the eigenvectors of non-interacting *G* such that only diagonal elements are retained, whereas, the number of eigenmodes is truncated in *sc-GW-subW* calculations. A systematic analysis of the results obtained from the above techniques is presented. The differences in the quasiparticle bandgap between the approximated and the full matrix *sc-GW* approaches are mostly less than 1.7% that validates such widely adopted approximations, and also shows how such low-ranking approximation can be used to include higher-order terms like the vertex correction.

`

†ysingh@lbl.gov
‡lwwang@lbl.gov

## 1. Introduction

The density functional theory (DFT) in the Kohn-Sham (KS) formulation [1,2] is an effective one-particle formalism and is well known for describing ground state properties of a wide range of materials. Despite the tremendous success of DFT, the bandgap problem using local density approximation (LDA) is well known. KS eigenvalues cannot be used to explain observed photoemission spectroscopy and optical absorption. However, such eigenvalues and eigenstates are generally used as a starting point for many-body excited-state calculations. [3,4] One of the several approaches in many-body perturbation theory is $GW$ approximation ($G$ is the Green's function, $W$ denotes the screened Coulomb interaction) which has now become a gold standard to describe quasiparticle and spectral properties of molecules and bulk materials. A single shot version of $GW$ ($G_0W_0$ approximation) has been quite successful in predicting experimental bandgaps of moderately correlated extended systems mainly because of the possible cancellation between self-consistency and vertex corrections. [5–8] The major shortcoming of the $G_0W_0$ method is its dependency on the choice of exchange-correlation (*xc*) functional (e.g. LDA or GGA or hybrid HSE or PBE0 etc.) that makes it less versatile, and often leads to different results in the literature. Moreover, the lack of self-consistency in $G$ leads to the violation of the conservation of particle number and total energy. [9–11] Attempt to include self-consistency in $G$ while keeping $W$ fixed ($GW_0$ approximation) has been explored showing the improved relative position of the occupied bands but failed to reproduce accurate quasiparticle energies. [5]

Fully self-consistent $GW$ schemes that update both $G$ and $W$ functions (e.g. *scGW* and *QSGW* methods, note the *scGW* notation should not be confused with *sc-GW*, which is used to indicate our method throughout the manuscript) have also been developed, tested and are supposed to be independent of the starting point. [12–19]. In the *QSGW* approach, for each iteration, $\omega$ independent non-interacting effective nonlocal potential is evaluated from the self-energy $\Sigma(i\omega)$ given by, $\sum_{i,j}\frac{1}{2}|\psi_i\rangle\{Re[\Sigma(\epsilon_i)]_{ij} + Re\,[\Sigma(\epsilon_j)]_{ij}\}\langle\psi_j|$ of the previous iteration step. This allows the use of non-interacting $G_0$ (Eq. (1)) in the

next iteration of *QSGW* calculation. [14,20]. The accuracy of the *QSGW* method might rely on the cancellation of quasiparticle approximation and neglect of attractive electron-hole interaction (vertex terms) contribution to self-energy and polarizability. However, in the *scGW* method of Shishkin and Kresse, in each iteration energy eigenvalues are updated in the Green's function as well as in the dielectric matrix of the screened potential, whereas wavefunctions are kept fixed.[12] In recent work, Grumet *et. al*. proposed an *scGW* method that utilizes imaginary frequency or time axis to calculate the correlated self-energy from $G_0W_0$ in the first step following which interacting Green's function is evaluated using the Dyson equation in the Hartree-Fock canonical-orbital basis.[17] It is important to mention here that both the above *scGW* approaches are within the framework of projector-augmented-wave (PAW) methodology. However, it is found that methods like *QSGW* or *scGW* systematically overestimate the bandgap with missing plasmonic satellites in the spectral function, primarily due to the underestimation of bulk dielectric constant $(\varepsilon)$. [12,17,19] Also for homogenous electron gas, Holm and von Barth compared $G_0W_0$ with an *scGW* method claiming self-consistency not only worsens the band structure results but it also broadens satellite structure making it featureless, thus, contradicting experimental findings. [21] To improve the quasiparticle and spectral properties of solids or molecules within full self-consistency, several attempts are being made to include computationally demanding vertex function that can be identified as the exchange of the RPA bubble diagrams encountered in polarizability calculation. [13,22–26] Such exchange diagrams can weaken the electron-electron interaction, thus effectively increase the dielectric constant. However, full implementation of the vertex correction based on the Feynman diagram will be extremely expensive, especially if such a term is represented as a full multi-dimensional tensor. It is thus necessary to seek for low-ranking approximation. [27] The other important effects that could contribute significantly to the quasiparticle and spectral properties of real materials are electron-phonon (e-ph) and spin-orbit couplings (SOC) which are there in an experiment but not accounted for in the regular *GW* calculations. [28–37] The contribution of e-ph coupling can be included in the *GW* self-energy according to Allen-Heine-Cardona theory using second-order perturbation theory within the adiabatic and harmonic approximations. [38–40] The inclusion of SOC becomes important when the system is magnetic or contains heavy earth metals. [33–36]

Most of the $GW$ codes use the following expression of the non-interacting Green's function,

$$G_0(\boldsymbol{r_1}, \boldsymbol{r_2}, \omega) = \sum_j \frac{\psi_j(\boldsymbol{r_1})\psi_j^*(\boldsymbol{r_2})}{\omega - [\epsilon_j - \mu + i\delta \, \mathrm{sgn}(\mu - \epsilon_j)]} \qquad (1)$$

where, $\epsilon_j$ is the eigenenergy corresponding to the single-particle eigenstate $\psi_j(\boldsymbol{r})$ which are orthonormal to each other, $\mu$ corresponds to the Fermi energy, and a small imaginary $\delta$ is required for the convergence of the Fourier transform to frequency space. The above expression is exact for non-interacting systems; however, it is only an approximation if used for interacting system Green's function. We do note that for an interacting system Green's function, there is a Lehmann representation identical to Eq. (1) [45], but with $\psi_j(\boldsymbol{r})$ being replaced by non-orthonormal quasiparticle amplitudes $f_j(\boldsymbol{r_1})$, and each $j$ representing one many-body excited state that can be reformulated to represent a quasi-particle orbital. It is difficult to use such $f_j(\boldsymbol{r_1})$ to do actual calculations, we will thus stay with the conventional usage of orthonormal quasiparticle state $\psi_j(\boldsymbol{r})$ in Eq.(1). If $\omega$ dependent $\psi_j(r, \omega)$ are used, then one can always represent an interacting $G$ in a diagonal form as in Eq. (1) (albeit a left and right wavefunctions are used, and $\omega$ dependent $\epsilon_j(\omega)$ is also used). However, in practice, $\omega$ independent $\psi_j(r)$ are always used. [12–14,20] In that case, the diagonal expression of Eq. (1) will constitute an approximation for the interacting $G$ which satisfies the Dyson equation. The $G(\omega)$ which satisfies the Dyson equation is a general matrix, with different eigenvectors at different $\omega$ values. As a result, it cannot be expressed as in Eq. (1) with a single set of $\omega$ independent $\psi_j(r)$. On the other hand, the Dyson equation can be considered as a variational minimum solution of Klein's total energy expression, and it follows many conservation laws including charge and particle number. [9,10,41] The true solution of Dyson's equation for $G$ will have a full matrix representation,

$$G^{-1}(i\omega) = i\omega + \mu - H_0 - \Sigma(i\omega) \qquad (2)$$

where $H_0$ is the single-particle Hamiltonian and $\Sigma(i\omega)$ is the self-energy. Both the $G(i\omega)$ and $\Sigma(i\omega)$ are full matrix depending on $\omega$. In this work, we first perform a plane-wave based full matrix self-consistent

*GW* (*sc-GW* method) calculation on a few semiconducting systems without resorting to diagonal *G* and low-rank *W* approximations. [19] To avoid singularity on the real axis, we have followed Rojas *et al*. "space-time" method to solve the above $G$ along the imaginary $i\omega + \mu$ axis in the complex $\omega$ plane. [42] On comparing our *sc-GW* method with Kresse's *scGW* scheme [17], we consider fully norm-conserving semi-core pseudopotentials in contrast with their projected augmented wave (PAW) pseudopotential. More importantly, we have kept a full matrix form for $G$ and $W$, which allows us to test the accuracies of diagonal $G$ and low-rank $W$ approximations. To deal with the head divergence at $\boldsymbol{k}$=0, we adopt an interpolation technique for the bandstructure using an extremely dense $(126 \times 126 \times 126)$ interpolated $\boldsymbol{k}$-point grid for the integration. [19] Other works have shown even a much smaller $\boldsymbol{k}$-point grid should be enough. For example, in the *scGW* work [17], $\boldsymbol{k}$-point grids from $(2 \times 2 \times 2)$ to $(6 \times 6 \times 6)$ were tested, and it was found the $(6 \times 6 \times 6)$ $\boldsymbol{k}$-point grid is fully converged. The results obtained from our *sc-GW* method considering full *G* and *W* matrices should be "exact" $GW$ results in a sense, as long as the $\boldsymbol{k}$-point grid is sufficient, plane-wave cutoff, and $\omega$ integration are converged. Within the present non-relativistic *GW* approach and besides convergence issues, the only source of approximation is the pseudopotential for which we use norm-conserving pseudopotentials with semi-core levels. [43,44] Our results can act as a benchmark to test the accuracies of the methods which employ diagonal *G* or low-ranking *W* approximations. This can also serve as a preparation for the vertex implementation.

Storing the full *G* matrix drastically increases computational cost. On the other hand, the implementation of diagonal-*G* (*sc-GW-diagG*) and low-rank subspace-*W* (*sc-GW-subW*) approximations can significantly reduce computational time and memory as it is done in most of the literature. The main goal of our current work is to test the accuracies of such approximations. In our *sc-GW-diagG* method, we have approximated the interacting $G(i\omega)$ in a diagonal form as,

$$G(r_1, r_2, i\omega) = \sum_j \psi_j(r_1)\psi_j^*(r_2)f_j(i\omega) \tag{3}$$

Note approximation comes because$\psi_j(r)$ is $\omega$ independent. However, during self-consistent interactions, $\psi_j(r)$ will be updated, and $f_j(\omega)$ are independent functions for different $j$ (alternatively, one can also use the denominator of Eq. (1) but with $\omega$ dependent $\epsilon_j(\omega)$). In the *sc-GW-subW* approach, we truncate the number of its eigenmodes to represent the $W$ function. In the literature, there are many works on how to reduce the rank of the inverse dielectric matrix $\varepsilon^{-1}$, from the early works of directly using a small plane wave cutoff, to recent works of using a limited number of dielectric eigenmodes by solving Sternheimer-equation (as will be discussed later). Here, in our fully *sc-GW* calculation, we use a plane wave cutoff $E_{cut2}$ to represent $W$ larger than the cutoff $E_{cut}$ used to represent $G$ (ideally $E_{cut2}$ should be $4E_{cut}$ since $W$ is a second-order function of $G$, much like the charge density in plane-wave density functional theory calculation. In practice, we have used $E_{cut2}$=$2E_{cut}$. Instead of using eigenmodes to represent matrix $\varepsilon^{-1}$ , here we use eigenmodes to represent $W$, so it can be useful to make the multi-variable integral separable in future many-body perturbation calculations (e.g., to evaluate the vertex term). To investigate our proposed methods, we consider bulk AlAs, AlP, GaP, and ZnS as the prototype systems. A close agreement between the quasiparticle bandgap and dielectric constant values obtained from the above cost-effective approximations with the fully *sc-GW* method validate these widely used approximations in the literature and opens up the possibility to include vertex corrections in the future.

The rest of the paper is organized as follows: In the next section, we briefly present the description of our *sc-GW* methodology and the numerical techniques used for calculations. Then, we describe the framework of our *sc-GW-diagG* and our *sc-GW-subW* approaches. This is followed by the analysis of the results and comparison with other calculations and experiments in section 3 and in section 4 we present our conclusions.

## 2. Methodology

Our fully *sc-GW* scheme is based on the "space-time" method by Rojas *et al.*, where, the $G$ is solved along the imaginary axis $i\omega + \mu$ in the complex $\omega$ plane to avoid singularity on the real axis. [42] The $G$ for a periodic system is derived from the Dyson equation,

$$G^{-1}(\boldsymbol{k}, i\omega) = i\omega + \mu - H(\boldsymbol{k}) - \Sigma(\boldsymbol{k}, i\omega), \quad (4)$$

such that for each wave vector $\boldsymbol{k}$ in the first Brillouin zone (BZ) and frequency $i\omega$, $G$, $H$, and $\Sigma$ are the matrices either in real space ($\boldsymbol{r}$ index) or reciprocal space ($\boldsymbol{q}$ index). The single-particle non-interacting Hamiltonian is given as, $H(\boldsymbol{k}) = -\frac{1}{2}(\nabla_r + i\boldsymbol{k})^2 + V(r) + \boldsymbol{\Sigma}_l|\phi_{l,\boldsymbol{k}}\rangle\langle\phi_{l,\boldsymbol{k}}|$ , where $-\frac{1}{2}\nabla_{\mathbf{r}}^2$ and $\boldsymbol{\Sigma}_l|\phi_{l,\boldsymbol{k}}\rangle\langle\phi_{l,\boldsymbol{k}}|$ denotes the kinetic energy the non-local pseudopotential projection operators, respectively. The one-electron potential $V(\boldsymbol{r}) = \boldsymbol{\Sigma}_R v_{at}(\boldsymbol{r}-\boldsymbol{R}) + \int \frac{\rho(r')}{|\boldsymbol{r}-\boldsymbol{r}'|} d^3\boldsymbol{r}'$ holds the local part of the atomic pseudopotential $v_{at}$ , Hartree potential $(v_H)$ with electron charge density calculated as $\rho(r) = -iG(\boldsymbol{r},\boldsymbol{r},i\tau)|_{\tau\to 0^+}$. $\Sigma$ denotes the electron self-energy and contains all the exchange-correlation effects. The DFT exchange-correlation function is only used in the first step. We have simply used LDA functional as our final converged result does not depend on the initial inputs. The transformation of the above functions from $\boldsymbol{r}$ to $\boldsymbol{q}$ space or vice versa is achieved by the following general expression,

$$X(\boldsymbol{q_1},\boldsymbol{q_2},\boldsymbol{k},z) = \frac{1}{\Omega}\int X(\boldsymbol{r_1},\boldsymbol{r_2},\boldsymbol{k},z)e^{i(q_1)r_1}e^{-i(q_2)r_2}d^3\boldsymbol{r_1}d^3\boldsymbol{r_2} \tag{5a}$$

$$\mathrm{X}(\boldsymbol{r_1},\boldsymbol{r_2},\boldsymbol{k},z) = \frac{1}{\Omega}\Sigma_{\boldsymbol{q_1 q_2}} X(\boldsymbol{q_1},\boldsymbol{q_2},k,z)e^{-i(q_1)r_1}e^{i(q_2)r_2}, \tag{5b}$$

where $\Omega$ denotes the volume of the unit cell, $\boldsymbol{r}$ and $\boldsymbol{q}$ are the lattice vectors in real and reciprocal space, respectively, $z$ is either $i\omega$ or $i\tau$. In the above equation, integration over $\boldsymbol{r_1}$ and $\boldsymbol{r_2}$ are done within a unit cell on a $N_r = 32\times32\times32$ numerical grid, while the summation over $\boldsymbol{q_1}$ and $\boldsymbol{q_2}$ are done within $E_{cut}$ plane wave vector sphere. The analytical expression of $G$ in $i\tau$ space is given by, [15]

$$G(\boldsymbol{q_1},\boldsymbol{q_2},k,i\tau) = \begin{cases} i\displaystyle\sum_{j,\epsilon_j<\mu} \psi_{j,k}(\boldsymbol{q_1})\psi^*_{j,k}(\boldsymbol{q_2})e^{\tau(\epsilon_j-\mu)}, \text{for } \tau>0 \\ -i\displaystyle\sum_{j,\epsilon_j>\mu} \psi_{j,k}(\boldsymbol{q_1})\psi^*_{j,k}(\boldsymbol{q_2})e^{\tau(\epsilon_j-\mu)}, \text{for } \tau<0. \end{cases} \tag{6}$$

Throughout the manuscript, we use $\tau' = i\tau$ and $\omega' = i\omega$ to represent the time and frequency variables, $\tau$ and $\omega$ are real numbers. This will make $\tau'$ and $\omega'$ on the imaginary axis. Note, $e^{i\omega.\tau} = e^{-i\omega'.\tau'}$, the Fourier transformation of the above the $G$ between $\tau$ and $\omega$ is given by,

$$G(\boldsymbol{q_1},\boldsymbol{q_2},k,i\tau) = \frac{i}{2\pi}\int_{-\infty}^{\infty} G(\boldsymbol{q_1},\boldsymbol{q_2},k,i\omega)\, e^{i\omega.\tau} d\omega \tag{7a}$$

$$G(\boldsymbol{q_1},\boldsymbol{q_2},k,i\omega) = -i\int_{-\infty}^{\infty} G(\boldsymbol{q_1},\boldsymbol{q_2},k,i\tau)\, e^{-i\omega.\tau} d\tau \tag{7b}$$

Computationally, this is done with discretized Fast Fourier Transformation (FFT). The self-energy operator $\Sigma$ is evaluated from $G$ and $W$ in the real space and time domain to avoid time-consuming convolution in the frequency space. More specifically, Hedin obtained a simple low-order expression of self-energy as the product of $G$ and dynamical screened-coulomb interaction ($W$), so-called the $GW$ approximation. [45]

$$\Sigma(\boldsymbol{r_1},\boldsymbol{r_2},\boldsymbol{k},i\tau) = i\sum_{\boldsymbol{k_2}} G(\boldsymbol{r_1},\boldsymbol{r_2},\boldsymbol{k}-\boldsymbol{k_2},i\tau)W(\boldsymbol{r_1},\boldsymbol{r_2},\boldsymbol{k_2},i\tau)\, w_{\boldsymbol{k_2}}. \tag{8}$$

Here, $w_{\boldsymbol{k}}$ corresponds to the weights associated with the $\boldsymbol{k}$-points. The expression of the $W$ function in the reciprocal space for a given $i\omega$ reads as

$$W(\boldsymbol{q_1},\boldsymbol{q_2},\boldsymbol{k},i\omega) = \frac{4\pi}{|\boldsymbol{q_1}+\boldsymbol{k}||\boldsymbol{q_2}+\boldsymbol{k}|}\varepsilon^{-1}(\boldsymbol{q_1},\boldsymbol{q_2},\boldsymbol{k},i\omega), \tag{9}$$

with factor $\frac{4\pi}{|\boldsymbol{q_1}+\boldsymbol{k}||\boldsymbol{q_2}+\boldsymbol{k}|}$ representing the symmetrized Fourier transformed bare Coulomb interaction and the dielectric function ($\varepsilon$) expressed in terms of irreducible polarizability ($\chi$) which is further obtained as the product of $G$ matrices from different $\boldsymbol{k}$ vectors,

$$\varepsilon(\boldsymbol{q_1},\boldsymbol{q_2},\boldsymbol{k},i\omega) = \delta_{\boldsymbol{q_1}\boldsymbol{q_2}} - \chi(\boldsymbol{q_1},\boldsymbol{q_2},\boldsymbol{k},i\omega)\frac{4\pi}{|\boldsymbol{q_1}+\boldsymbol{k}||\boldsymbol{q_2}+\boldsymbol{k}|}. \tag{10}$$

$$\chi(\boldsymbol{r_1},\boldsymbol{r_2},\boldsymbol{k},i\tau) = -i\sum_{\boldsymbol{k_2}} G(\boldsymbol{r_1},\boldsymbol{r_2},\boldsymbol{k}+\boldsymbol{k_2},i\tau)G(\boldsymbol{r_2},\boldsymbol{r_1},\boldsymbol{k_2},-i\tau)\, w_{\boldsymbol{k_2}} \tag{11}$$

It is to be noted that self-energy given by Eq. (8) has a discrete summation over $\boldsymbol{k}$ and will give divergent results at $\boldsymbol{k}$=0 because of the divergence of screened-interaction $W$ at $\boldsymbol{k}$=0 as shown in Eq. (9). This divergence problem is handled by following a similar technique used for unscreened Fock exchange term calculation by Gygi and Baldereschi [46]. Here, a reference term that also diverges at $\boldsymbol{k}$=0 is added and subtracted from the right-hand side of Eq. (8) and a dense grid in the 1$^{st}$ BZ is used to avoid divergence problem in the second term. A more detailed description is given in reference [19]. In this work, instead of using Matsubara time and frequency mesh at an artificial temperature [47,48], we carry out integrations in $\omega$ and $\tau$ space on a discrete exponential numerical grid as explained in our previous publication. [15] The evaluation of the dielectric function in Eq. (10) is the trickiest part of the *sc-GW* calculation because of the **Γ** point divergence problem at $\boldsymbol{k}$=0 in periodic systems. To escape this singularity at small $\boldsymbol{k}$ values, we expand the head ($\boldsymbol{q_1} = \boldsymbol{q_2} = 0$) and the wing ($\boldsymbol{q_1} = 0$ or $\boldsymbol{q_2} = 0$) of the $\chi(\boldsymbol{q_1}, \boldsymbol{q_2}, \boldsymbol{k}, i\omega)$ matrix as a function of $\boldsymbol{k}$. We follow a technique similar to Hybersten and Louie's [49] single-particle expression of the $\chi$ function that uses Alder-Wiser formulation [50,51] to deduce "head" and "wing" expansion of the polarizability $\chi$ in the limits $\boldsymbol{k} \to 0$. To describe the head case, we expand the polarizability in the limits $\boldsymbol{k} \to 0$ as [19],

$$\begin{aligned}
&\chi(\boldsymbol{q_1} = 0, \boldsymbol{q_2} = 0, \boldsymbol{k}, i\tau) \\
&= \chi(\boldsymbol{q_1} = 0, \boldsymbol{q_2} = 0, \boldsymbol{k} = 0, i\tau) + \sum_{\alpha,\beta} \chi^{(2)}_{\boldsymbol{\alpha},\boldsymbol{\beta}}(\boldsymbol{q_1} = 0, \boldsymbol{q_2} = 0, \boldsymbol{k} = 0, i\tau)\boldsymbol{k_\alpha}\boldsymbol{k_\beta} \\
&= \sum_{\alpha,\beta} \chi^{(2)}_{\boldsymbol{\alpha},\boldsymbol{\beta}}(\boldsymbol{q_1} = 0, \boldsymbol{q_2} = 0, \boldsymbol{k} = \boldsymbol{0}, i\tau)\boldsymbol{k_\alpha}\boldsymbol{k_\beta}
\end{aligned} \tag{12}$$

where, $\boldsymbol{k_\alpha}$ or $\boldsymbol{k_\beta}$ ($\boldsymbol{\alpha}$ or $\boldsymbol{\beta} = x, y$ or $z$) are the directional components of $\boldsymbol{k}$ approaching **Γ** point ($\boldsymbol{k} = 0$). The first term in the above expression is zero and to get term $\chi^{(2)}_{\boldsymbol{\alpha},\boldsymbol{\beta}}(\boldsymbol{q_1} = 0, \boldsymbol{q_2} = 0, \boldsymbol{k} = \boldsymbol{0}, i\tau)$, we define a middle step term $\chi^{(0)}_{\boldsymbol{\alpha},\boldsymbol{\beta}}(\boldsymbol{q_1} = 0, \boldsymbol{q_2} = 0, \boldsymbol{k} = 0, i\tau)$ based on the Eq. (11) that reads as

$$\begin{aligned}
&\chi^{(0)}_{\boldsymbol{\alpha},\boldsymbol{\beta}}(\boldsymbol{q_1} = 0, \boldsymbol{q_2} = 0, \boldsymbol{k} = 0, i\tau) \\
&= -i \sum_{k_2} \int \Big[ \{\boldsymbol{\nabla}^{\boldsymbol{\alpha}}_{\boldsymbol{k_2}} H(\boldsymbol{r_1}, \boldsymbol{k_2})\} G(\boldsymbol{r_1}, \boldsymbol{r_2}, \boldsymbol{k_2}, i\tau) \\
&\times \{\boldsymbol{\nabla}^{\boldsymbol{\beta}}_{\boldsymbol{k_2}} H(\boldsymbol{r_2}, \boldsymbol{k_2})\} G(\boldsymbol{r_2}, \boldsymbol{r_1}, \boldsymbol{k_2}, -i\tau) w_{\boldsymbol{k_2}} \Big] d^3\boldsymbol{r_1} d^3\boldsymbol{r_2}
\end{aligned} \tag{13}$$

Here,

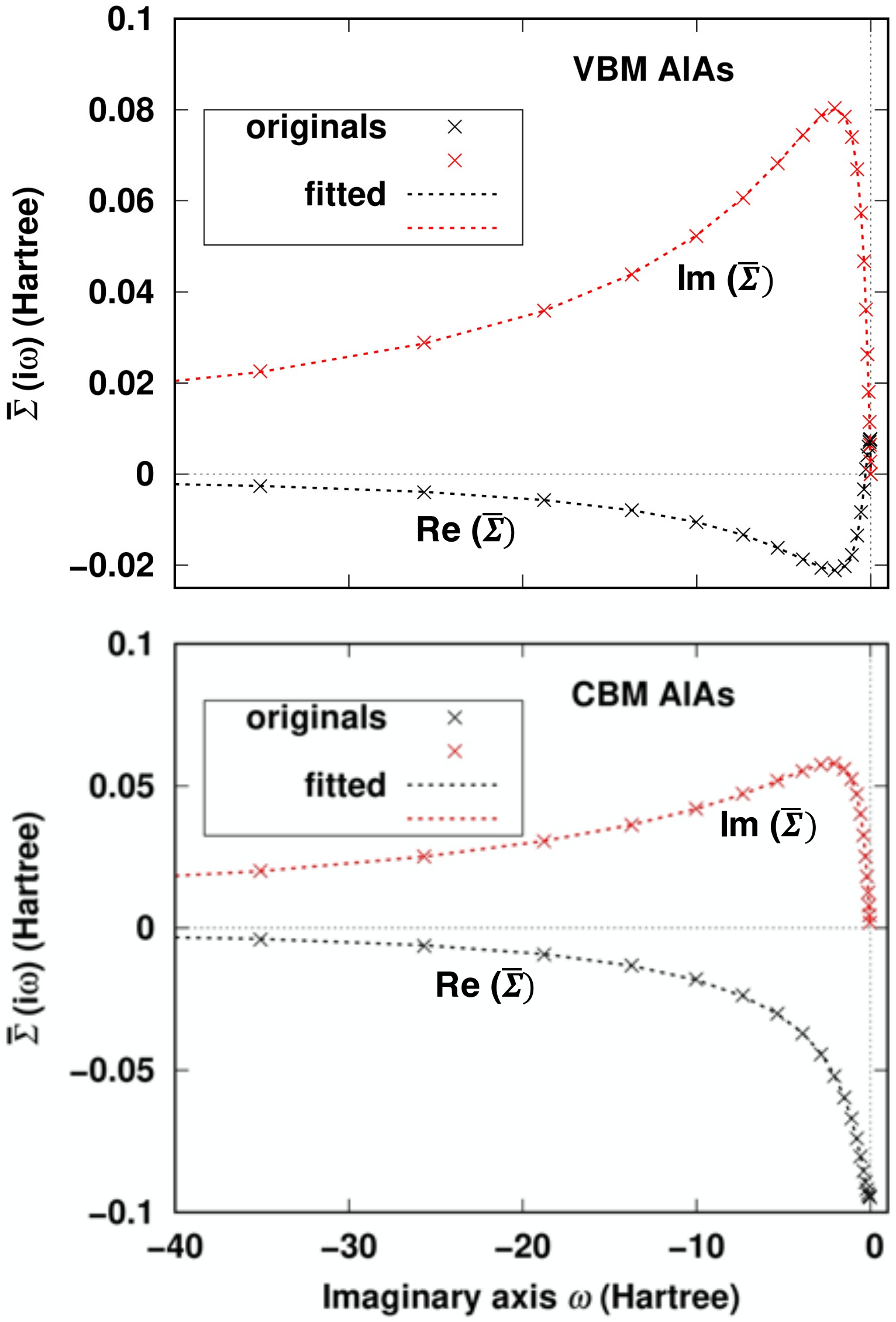

**Figure 1.** The expectation value $\bar{\Sigma}_{jk}(i\omega) = \langle\psi_{jk}(i\omega = 0)|\Sigma(\boldsymbol{k}, i\omega)|\psi_{jk}(i\omega = 0)\rangle$ of AlAs for (a) $j$=13 (VBM) and (b) $j$=14 (CBM) at the **Γ** point ($\boldsymbol{k}$=0) on the imaginary axis.

$\boldsymbol{\nabla}^{\alpha}_{\boldsymbol{k_2}} H(\boldsymbol{r_1}, \boldsymbol{k_2})$ is the derivative of single-particle Hamiltonian ($H(\boldsymbol{r}, \boldsymbol{k}) = e^{-i\boldsymbol{k}.\boldsymbol{r}} H(\boldsymbol{r}) e^{i\boldsymbol{k}.\boldsymbol{r}}$) with respect to $\boldsymbol{k_2}$ along the $\boldsymbol{\alpha}$ direction and in the calculation, it is written as a matrix to represent non-local term ($\boldsymbol{k_2}$ belongs to the original $\boldsymbol{k}$-grid in the first BZ). In the non-interacting formalism, $\chi^{(2)}_{\boldsymbol{\alpha},\boldsymbol{\beta}}(\boldsymbol{q_1} = 0, \boldsymbol{q_2} = 0, \boldsymbol{k} = 0, i\tau)$ can be equated to $\chi^{(0)}_{\boldsymbol{\alpha},\boldsymbol{\beta}}(\boldsymbol{q_1} = 0, \boldsymbol{q_2} = 0, \boldsymbol{k} = 0, i\tau)$ divided by eigenenergy square term which is obtained by a double integration of $\tau$ in the form (for $\tau > 0$), [51,52]

$$\chi^{(2)}_{\boldsymbol{\alpha},\boldsymbol{\beta}}(\boldsymbol{q_1}=0,\boldsymbol{q_2}=0,\boldsymbol{k}=0,i\tau)=-\int_{\infty}^{\tau}\left[-\int_{\infty}^{\tau'}\chi^{(0)}_{\boldsymbol{\alpha},\boldsymbol{\beta}}(\boldsymbol{q_1}=0,\boldsymbol{q_2}=0,\boldsymbol{k}=0,i\tau'')d\tau''\right]d\tau' \tag{14}$$

$$=\sum_{k_2}-i\left\{-\int_{\infty}^{\tau}\left[-\int_{\infty}^{\tau'}[\int\left[\boldsymbol{\nabla}^{\boldsymbol{\alpha}}_{\boldsymbol{k_2}}H(\boldsymbol{r_1},\boldsymbol{k_2})G(\boldsymbol{r_1},\boldsymbol{r_2},\boldsymbol{k}+\boldsymbol{k_2},i\tau'')\right.\right.\right.$$

$$\left.\left.\left.\times\boldsymbol{\nabla}^{\boldsymbol{\beta}}_{\boldsymbol{k_2}}H(\boldsymbol{r_2},\boldsymbol{k_2})G(\boldsymbol{r_2},\boldsymbol{r_1},\boldsymbol{k_2},-i\tau'')w_{k_2}\right]d^3\boldsymbol{r_1}d^3\boldsymbol{r_2}]\,d\tau''\right]d\tau'\right\}.$$

For $\tau<0,$ the integrals should start from minus infinity. The approximated head expression in Eq. (14) is rigorous only for non-interacting $G$ but should capture the main contribution of the $\boldsymbol{k}$ expansion. [19] Eq. (12) is used to evaluate $\chi(\boldsymbol{q_1}=0,\boldsymbol{q_2}=0,\boldsymbol{k},i\tau)$ for any $\boldsymbol{k}$ points near $\boldsymbol{k}\to 0$. The expression for the "wings" part is evaluated in a similar fashion using one derivative only as (details are given in Appendix B of Ref [19]),

$$\chi(\boldsymbol{q_1}=0,\boldsymbol{q_2},\boldsymbol{k},i\tau)=-\sum_{\boldsymbol{\alpha}}\int_{\infty}^{\tau}\chi^{(0)}_{\boldsymbol{\alpha}}(\boldsymbol{q_1}=0,\boldsymbol{q_2},\boldsymbol{k},i\tau')\boldsymbol{k_\alpha}\,d\tau', \tag{15}$$

where,

$$\chi^{(0)}_{\boldsymbol{\alpha}}(\boldsymbol{q_1}=0,\boldsymbol{q_2},\boldsymbol{k},i\tau)=-i\sum_{\boldsymbol{k_2}}\int\left[\boldsymbol{\nabla}^{\boldsymbol{\alpha}}_{\boldsymbol{k_2}}H(\boldsymbol{r_1},\boldsymbol{k_2})G(\boldsymbol{r_1},\boldsymbol{r_2},\boldsymbol{k}+\right.$$

$$\left.\boldsymbol{k_2},i\tau)G(\boldsymbol{r_2},\boldsymbol{r_1},\boldsymbol{k_2},-i\tau)w_{\boldsymbol{k_2}}e^{-\boldsymbol{i}q_2r_2}\right]d^3\boldsymbol{r_1}d^3\boldsymbol{r_2}.$$

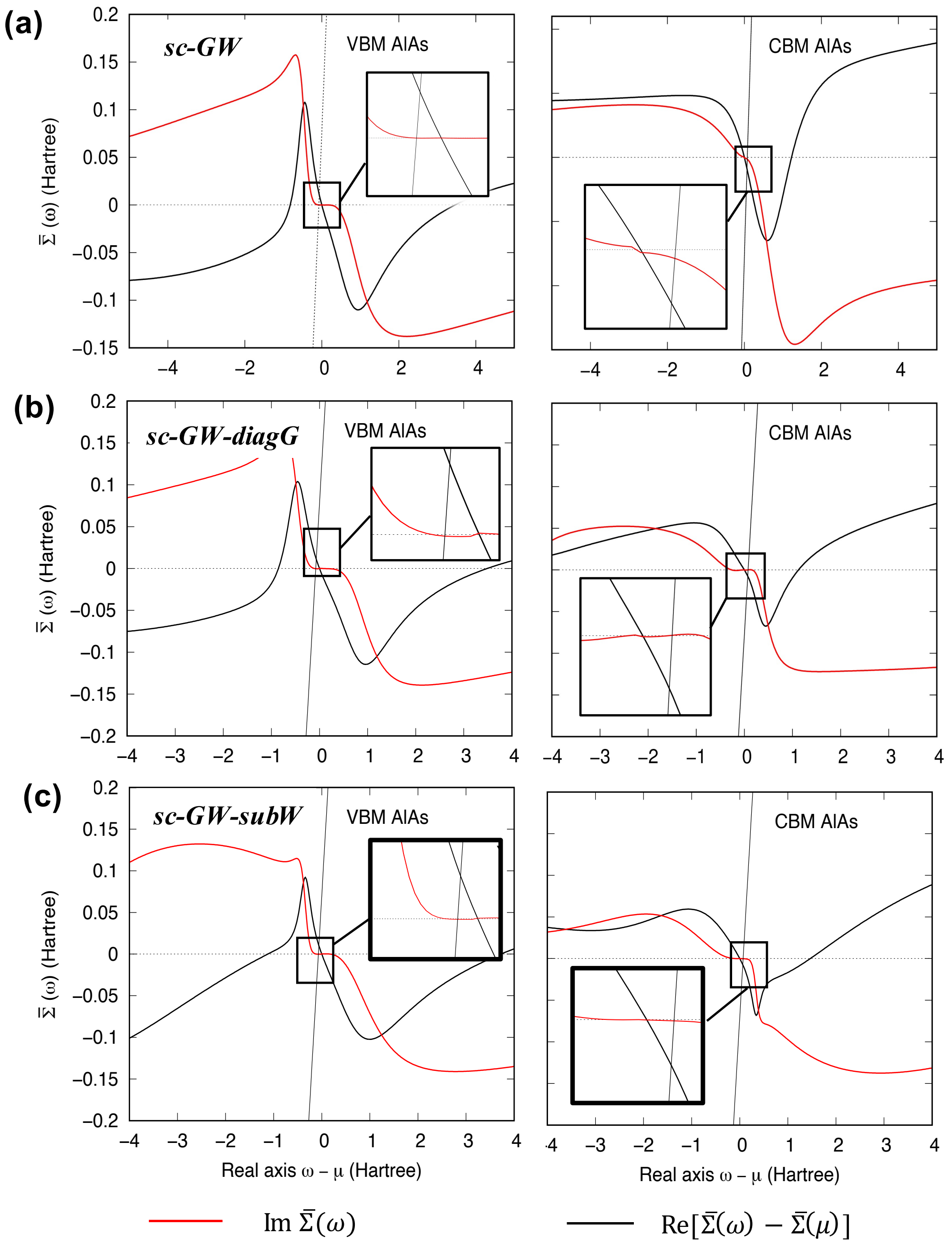

**Figure 2.** The expectation value $\bar{\Sigma}_{\boldsymbol{jk}}(\omega)$ on a real frequency axis obtained from the analytical extension of $\bar{\Sigma}_{\boldsymbol{jk}}(i\omega)$ for *j*=13 (VBM) and *j*=14 (CBM) states of AlAs at the $\boldsymbol{\Gamma}$ point ($\boldsymbol{k}$=0) obtained from the (a) *sc-GW* (b) *sc-GW-diagG*, and (c) *sc-GW-subW* methods.

Once we determine $\chi$, the expression for the dielectric function can be approximated to:

$$\varepsilon(0,0,\boldsymbol{k},i\omega) = 1 - 4\pi \sum_{\boldsymbol{\alpha},\boldsymbol{\beta}} \frac{k_\alpha k_\beta}{\boldsymbol{k}^2} \sum_{k_2} \bar{\chi}_{\boldsymbol{\alpha},\boldsymbol{\beta}}(0,0,\boldsymbol{k_2},i\omega)\, w_{\boldsymbol{k_2}}. \tag{16}$$

Here $\bar{\chi}_{\boldsymbol{\alpha},\boldsymbol{\beta}}(0,0,\boldsymbol{k_2},i\omega)$ denotes the contribution of each $\boldsymbol{k_2}$ point in the $\boldsymbol{k_2}$ summation of Eq. (14) and (15) (the values within the curly brackets). To compare with experiments, one can obtain the macroscopic dielectric constant ($\varepsilon$) for cubic material (which is $\boldsymbol{k} \to \boldsymbol{0}$ orientation independent) as $1/\varepsilon^{-1}(\boldsymbol{q_1} = 0, \boldsymbol{q_2} = 0, \boldsymbol{k} = 0, i\omega = 0)$ including the local field effects which are important to predict the correct quasiparticle spectrum. [53]

**Diagonal *G* and subspace *W* approximations**: Most of the *GW* methods start with the non-interacting $G_0$ (Eq. (1)) and during self-consistency, the contribution from $\Sigma$ is included via updating $\epsilon_i$ and/or $\psi_i$ iteratively. In another approach, the frequency dependency of $\Sigma(i\omega)$ in the Dyson equation (Eq. (2)) is approximated by a linear function that introduces a renormalization factor *Z* which transform Eq. (2) into an approximate quasiparticle equation. [12,14] However, in our *sc-GW-diagG* method, we seek for a general diagonal approximation of $G$ for a non-zero $\omega$ in each iteration. The full $G$ matrix at $i\omega \neq 0$ derived from the Dyson equation is expanded in terms of the eigenvectors $\psi_{j,\boldsymbol{k}}(\boldsymbol{q})$ of $G(\boldsymbol{q_1}, \boldsymbol{q_2}, \boldsymbol{k}, i\omega = 0)$ for each $\boldsymbol{k}$, and the off-diagonal elements are ignored, such that,

$$G(\boldsymbol{q_1}, \boldsymbol{q_2}, \boldsymbol{k}, i\omega) = \sum_j \psi_{j,\boldsymbol{k}}(\boldsymbol{q_1})\psi^*_{j,\boldsymbol{k}}(\boldsymbol{q_2})\, f_{j,\boldsymbol{k}}(i\omega). \tag{17}$$

where, $f_{j,\boldsymbol{k}}(i\omega)$ contains the frequency dependency, which is obtained from $<\psi_{j,\boldsymbol{k}}|G(\boldsymbol{q_1}, \boldsymbol{q_2}, \boldsymbol{k}, i\omega)|\psi_{j,\boldsymbol{k}}>$. In our calculation, $\psi_{j,\boldsymbol{k}}(\boldsymbol{q})$ of the static $G(\boldsymbol{q_1}, \boldsymbol{q_2}, \boldsymbol{k}, i\omega = 0)$ were obtained in the first iteration which is then used to construct the matrix elements of dynamic $G(\boldsymbol{q_1}, \boldsymbol{q_2}, \boldsymbol{k}, i\omega)$ at each frequency point in subsequent iterations. A comparison of diagonal and non-diagonal components of the $G$ matrix in this new basis will be presented to highlight their significance.

The major challenge in $GW$ calculations lies in the full-frequency evaluation of the $\chi$ matrix that scales as $O(N^4)$ where N is proportional to the system size. Moreover, a huge memory requirement to store above matrices is the major issue to calculate the vertex function. In an attempt to speed up above computation, Wilson *et al.* [54,55], Nguyen *et al.* [56], and Pham *et al.* [57] used the eigendecomposition of static $\chi(\omega = 0)$ as a basis set to estimate $\chi(\omega \neq 0)$ by solving Sternheimer-equation within density functional perturbation theory (DFPT). This approach not only avoids the calculation of a large number of excited states but also avoids the inversion of the ε matrix as required by other explicit perturbation-based methods. Inspired from the above work and motivated from the properties of dielectric function [58–60], Ben et al. [61] came up with the idea of static subspace approximation where they bypass DFPT and consider a static basis set obtained from a low-rank approximation [27] of the symmetrized static dielectric function. The number of eigenvectors is selected based on a given eigenvalue threshold. This static basis set is used to evaluate the matrix elements of the polarizability function for the finite non-zero values of $\omega$. It's worth to mention here that, individual eigenvectors corresponding to $\varepsilon^{-1}(\omega)$ might vary considerably with $\omega$, but the subspace expanded by this basis set of eigenvectors (chosen from a given eigenvalue threshold) varies insignificantly with ω. [57,62–65] Our approach here is quite similar to these works. However, since the expression of self-energy is the direct product of $G$ and $W$ matrix, considering a direct low-ranking approximation to $W$ instead of $\varepsilon^{-1}$ would be more straightforward. In our *sc-GW-subW* method, we directly consider the full $W$ matrix and drastically reduce its dimension by a fixed number of eigenvectors. In this method, we deal with the dynamic part of $W$ by subtracting frequency-independent bare Coulomb interaction from the $W$ function with the remaining matrix called $W'$. Similar to the *sc-GW-diagG* method, we consider eigenfunctions ($\theta_{j,\boldsymbol{k}}(\boldsymbol{q})$) corresponding to the static $W'(\boldsymbol{q_1}, \boldsymbol{q_2}, \boldsymbol{k}, i\omega = 0)$ matrix as the basis set to evaluate $W'(\boldsymbol{q_1}, \boldsymbol{q_2}, \boldsymbol{k}, i\omega)$ at all the frequencies.

$$W'(\boldsymbol{q_1}, \boldsymbol{q_2}, \boldsymbol{k}, i\omega) = W(\boldsymbol{q_1}, \boldsymbol{q_2}, \boldsymbol{k}, i\omega) - \delta_{q_1,q_2} \frac{4\pi}{|\boldsymbol{q_1} + \boldsymbol{k}|^2} \tag{18}$$

$$W'(\boldsymbol{q_1}, \boldsymbol{q_2}, \boldsymbol{k}, i\omega) = \sum_{j_1 j_2} \theta_{j_1,\boldsymbol{k}}(\boldsymbol{q_1})\, V(j_1, j_2, \boldsymbol{k}, i\omega)\, \theta^*_{j_2,\boldsymbol{k}}(\boldsymbol{q_2}) \,. \tag{19}$$

Note, if we include all the mode $j_1$, $j_2$ in Eq. (19), it is exact. The approximation comes when we truncate the modes from the original tens of thousands to the order of 30 modes. This is thus a low-rank approximation. In the *sc-GW* framework, coupled Eq.'s (4), (8), (9) – (11) form a closed set of equations which are solved iteratively till the $G$ converges. It is worth to mention here that Hamiltonian $H(\boldsymbol{k}, i\omega) = H(\boldsymbol{k}) + \Sigma(\boldsymbol{k}, i\omega) = i\omega + \mu - G^{-1}(\boldsymbol{k}, i\omega)$ is Hermitian at $\omega = 0$, thus it can be diagonalized to yield the eigenstate wavefunctions. The starting $G$ function is constructed using LDA Kohn-Sham eigenvalues and eigenfunctions as inputs, therefore at the first iteration, our results are equivalent to the conventional $G_0W_0$ calculations.

## Technical Details

In our *sc-GW* scheme, we use a norm-conserving semi-core pseudopotential to have the correct exchange integral, thus avoiding systematic errors that are built-in pseudopotential-based *GW* schemes. [13,66] We adopted four prototype systems (AlAs, AlP, GaP, and ZnS) with cubic zinc blend structure with experimental lattice parameters as shown in Table S1 of the supporting information (SI). The electronic configuration for the valence electrons for Al, As, P, Ga, Zn, and S atoms are described as $3s^2 3p^1$, $3s^2 4s^2 3p^6 4p^3 3d^{10}$, $3s^2 3p^3$, $3s^2 4s^2 3p^6 4p^1 3d^{10}$, $3s^2 4s^2 3p^6 4d^{10}$, and $3s^2 3p^4$, respectively. As described in our previous work a large kinetic energy cutoff ($E_{cut}$: required to expand $G$ and $\Sigma$ functions) is required to get well-converged results owing to the highly localized semi-core states. [19] Such dependency on $E_{cut}$ can be seen in the LDA bandgaps which are compared at 75 Ry and a sufficiently large value of 300 Ry (see Table S1). For AlP, no changes in the bandgap value were observed due to $E_{cut}$. But for other systems, there could be a significant difference between the 75 Ry $E_{cut}$ results and 300 Ry $E_{cut}$ results. One can safely assume LDA bandgap calculated at 300 Ry is converged, but using $E_{cut}$=300 Ry for the *sc-GW* calculation can significantly increase the memory requirement and the computational cost.

It has been shown that plane wave truncation mildly affects the valence energy levels, thus, it can be treated as a perturbation to the Hamiltonian. [19] To counter this effect, the original, *s,* or *p,* or *d* potentials are tweaked by adding a small Gaussian function *g(r)* that reads as

$$g(\boldsymbol{r}) = \beta \mathrm{e}^{-\left[\frac{(\boldsymbol{r}-\boldsymbol{r}_{peak})}{\boldsymbol{r}_{cut}}\right]^2} \tag{20}$$

where, $\boldsymbol{r}$, $\boldsymbol{r}_{peak}$, and $\boldsymbol{r}_{cut}$ are the radius, position of the peak in the radial direction, and width of the Gaussian in the units of Bohr, respectively. We obtain the revised pseudopotentials that reproduce converged LDA

**Table I. Quasiparticle bandgaps ($E_{gap}$) in (eV) for bulk AlAs, AlP, GaP and ZnS systems obtained from LDA, $G_0W_0$, fully *sc-GW*, *sc-GW-diagG,* and *sc-GW-subW* methods at the Γ point and compared with other calculations and experiments. The $G_0W_0$ results in the references [12,17], [13], and [67,68] are obtained using PBE, HSE03, and LDA, respectively.**

| System | $E_{gap}$ (eV) This work | | | | | $E_{gap}$ (eV) Other | |
|---|---|---|---|---|---|---|---|
| | LDA | $G_0W_0$ | *sc-GW* | | | Calculations (Method) | Experiments |
| | | | *fully* | *diagG* | *subW* | | |
| AlAs | 1.44 | 2.50 | 3.65 | 3.70 | 3.67* | **3.58** [17], **3.35** [69] (*QPGW*)<br>**2.97** [13] ($G_0W_0$)<br>**3.73** [17] (*scGW*) | **3.13 (0.01)** [70] |
| AlP | 3.13 | 3.92 | 4.36 | 4.41 | 4.38 | **4.20** [13] ($G_0W_0$)<br>**5.01** [17] (*scGW*)<br>**4.74** [17] (*QPGW*) | **3.62(0.02)** [70] |
| GaP | 1.55 | 2.53 | 3.00 | 3.00 | 3.05 | **2.62** [17], **2.79** [67], **3.01** [68] ($G_0W_0$)<br>**3.17** [17] (*scGW*)<br>**3.05** [17] (*QPGW*) | **2.866** [70] |
| ZnS | 2.18 | 3.70 | 4.14 | 4.35 | 4.17 | **3.43** [17], **3.46** [13], **3.29** [12] ($G_0W_0$)<br>**4.68** [17], **4.15** [13], **3.86** [12] (*scGW*)<br>**4.27** [17] (*QPGW*) | **3.723(0.001)** [71]<br>**3.78** [72] |

**QP: Quasiparticle, *: only three iterations have been accomplished for this calculation. The value quoted here is the estimated converged value according to the iteration behavior of full *sc-GW*.**

bandgap values with $E_{cut}$ = 75Ry (see Table S1 in SI) by adjusting $\beta$, $\boldsymbol{r}_{peak}$, and $\boldsymbol{r}_{cut}$ values in the above function. The values of the above parameters to get slightly modified pseudopotentials for the considered systems are given in Table S2 of SI. A cutoff energy of $E_{cut}$ = 75 Ry yields approximately 2975 ($N_q$) plane wavevectors that are used to expand $G$ and Σ functions. However, a larger cutoff of $E_{cut2}$ = 150 Ry

**Table II. Comparison of the macroscopic dielectric constant ($\varepsilon$ which is technically $1/\varepsilon^{-1}(q_1 = 0, q_2 = 0, k = 0, \omega = 0)$ values obtained from the methods same as Table I. and compared with other calculations ($\varepsilon_{cal}$) and experimental high-frequency dielectric constant ($\varepsilon_{expt}$).**

| Systems | $\varepsilon$ (This work) | | | | $\varepsilon_{\infty}$ (Other) | |
|---|---|---|---|---|---|---|
| | $\varepsilon_{\text{LDA}}$ | $\varepsilon_{GW}$ | | | $\varepsilon_{cal}$ | $\varepsilon_{expt}$ |
| | | *full* | *diagG* | *subW* | | |
| **AlAs** | **10.32** | **3.01** | **2.93** | **2.99*** | ~ | **8.16** [67] |
| **AlP** | **9.31** | **4.59** | **4.44** | **4.56** | **7.11 (*scGW*)** [13] | **7.54** [73] |
| **GaP** | **14.15** | **5.79** | **5.72** | **5.75** | ~ | **9.06 (36)** [74] |
| **ZnS** | **4.99** | **3.41** | **3.31** | **3.36** | **5.15 (*scGW*)** [13] | **5.13** [75] |

***: only three iterations have been accomplished for this calculation. The value quoted here is the estimated converged value according to the iteration behavior of full *sc-GW*.**

is used to generate about 8609 ($N_{qL}$) plane waves to express $W$, $\varepsilon$, and $\chi$ matrices. The total number of grid points in real space is $N_r = 32 \times 32 \times 32 = 32768$ which is roughly ten times bigger than $N_q$.

The evaluation of $\varepsilon$ is the most time-consuming part of the algorithm whose convergence depends on the number of $\boldsymbol{k}$-point summation. [52,76] We consider full matrices without truncating the conduction band. In our previous publication, we observed that convergence of LDA dielectric constant is notoriously slow mainly due to the influence of large value at the $\boldsymbol{\Gamma}$ point and may require a denser $\boldsymbol{k}$-grid mesh. [19] This issue was handled in a two-fold way, first by employing a linear tetrahedron interpolation scheme that evaluates $\chi$ on a much denser $\boldsymbol{k}$ mesh $(126 \times 126 \times 126)$. Second, by assuming no significant changes in the shape of the $\chi$ function near $\boldsymbol{\Gamma}$ point (except for the magnitude) from LDA to $GW$ results. A much denser $(15 \times 15 \times 15)$ LDA $\boldsymbol{k}$-point grid than the $GW$ calculation $\boldsymbol{k}$-point grid (which is 6x6x6) is used to fix the shape of $\chi$ near the $\boldsymbol{\Gamma}$ point to be used for the linear interpolation scheme for the $\boldsymbol{k}$-point summation. These schemes assume both the accuracy of the results and fast convergence with $\boldsymbol{k}$-points.

The imaginary frequency and time axis to describe $G(i\omega)$ and $G(i\tau)$ matrices are divided into 100 and 80 exponential grid points, respectively. [15] The maximum and minimum of $\omega$ areis given as $3 \times 10^6$ and $2 \times 10^{-4}$ Hartree, respectively. The maximum and minimum values for $\tau$ are given as 200 and 0.01

Hartree$^{-1}$, respectively. The technicalities of the numerical Fourier transformation between $G(i\omega)$ and $G(i\tau)$ are given in our previous publication. [15]

Using crystal symmetry, we further reduce the computational cost by considering selected reduced $\boldsymbol{k}$-points that lie within the irreducible BZ. The self-consistency in our *sc-GW* iteration is said to be achieved till we get converged eigenenergy $\epsilon_{jk}(0)$ corresponding to $H'(k, i\omega)$ when $\omega = 0$ with a tolerance of 0.01 eV. To speed up the calculation we massively parallelized our *sc-GW* code and used 24000 processors on one of the largest supercomputers, Titan, at the Oak Ridge Leadership Computing Facility (OLCF). At the first level of parallelization, the $\boldsymbol{k}$-points are subdivided into process groups. In the second parallelization level within each processor group, $\boldsymbol{q_2}$ is divided into different processors. It takes about three hours for one iteration of *sc-GW* calculation.

## 3. Results and Discussions

We first show the importance of using the self-consistent $G$ matrix over non-interacting $G_0$ (Eq. (1)) in our calculations. The difference between $G$ and $G_0$ is the existence of $\Sigma$ function in $G$. Considering AlAs as an example, in Figure 1., we display the expectation value $\bar{\Sigma}_{j\boldsymbol{k}}(i\omega) = \langle\psi_{j\boldsymbol{k}}(i\omega = 0)|\Sigma(\boldsymbol{k}, i\omega)|\psi_{j\boldsymbol{k}}(i\omega = 0)\rangle$ for states *j*=13 and *j*=14 corresponding to valence band maximum (VBM) and conduction band minimum (CBM), respectively. Here, we approximated $|\psi_{j\boldsymbol{k}}\rangle \approx |\psi_{j\boldsymbol{k}}(i\omega = 0)\rangle$ because $|\psi_{j\boldsymbol{k}}\rangle$ is almost independent of $\omega$ at least for occupied and a few conduction band states near the gap. [15]. The magnitude of imaginary and real part of $\bar{\Sigma}$ vary significantly with $\omega$, thus, we cannot approximate $\bar{\Sigma}$ by its value at $\omega = 0$.

The calculation of quasiparticle energy and the spectral function which are often measured in the experiments require the knowledge of $G$ and $\Sigma$ functions in the real frequency domain. Therefore, we obtain $\bar{\Sigma}_{j\boldsymbol{k}}(\omega)$ on real axis by analytical extension of $\bar{\Sigma}_{j\boldsymbol{k}}(i\omega)$ as proposed by Rojas *et al.* [42]. The accuracy of the new analytical function is accessed from Figure 1. where the solid line fits well the original data. The solid line has an analytical form of $\sum_{l} \frac{a_l}{i\omega - z_l}$ with $a_l$ and $z_l$ being the fitting parameters and the number of terms considered is $l = 4$ . It can thus readily be extended to the real axis by simply changing $i\omega$ to $\omega$. The reliability of this procedure in obtaining $\bar{\Sigma}_{j\boldsymbol{k}}(\omega)$ for $\omega$ within 1 to 2 Hartree of $\mu$ is well tested. [15] Quasiparticle energies or poles of $G$ correspond to the $\omega$ solutions of $\langle \psi_{j\boldsymbol{k}}(\omega)|G^{-1}(\boldsymbol{k},\omega)|\psi_{j\boldsymbol{k}}(\omega)\rangle = 0$. If we restrict our solution to be on the real axis, i.e., $\omega$ to be a real number, then we have,

$$\omega + \mu = \epsilon_{j\boldsymbol{k}}(0) + \mathrm{Re}\big[\bar{\Sigma}_{j\boldsymbol{k}}(\omega) - \bar{\Sigma}_{j\boldsymbol{k}}(0)\big], \tag{21}$$

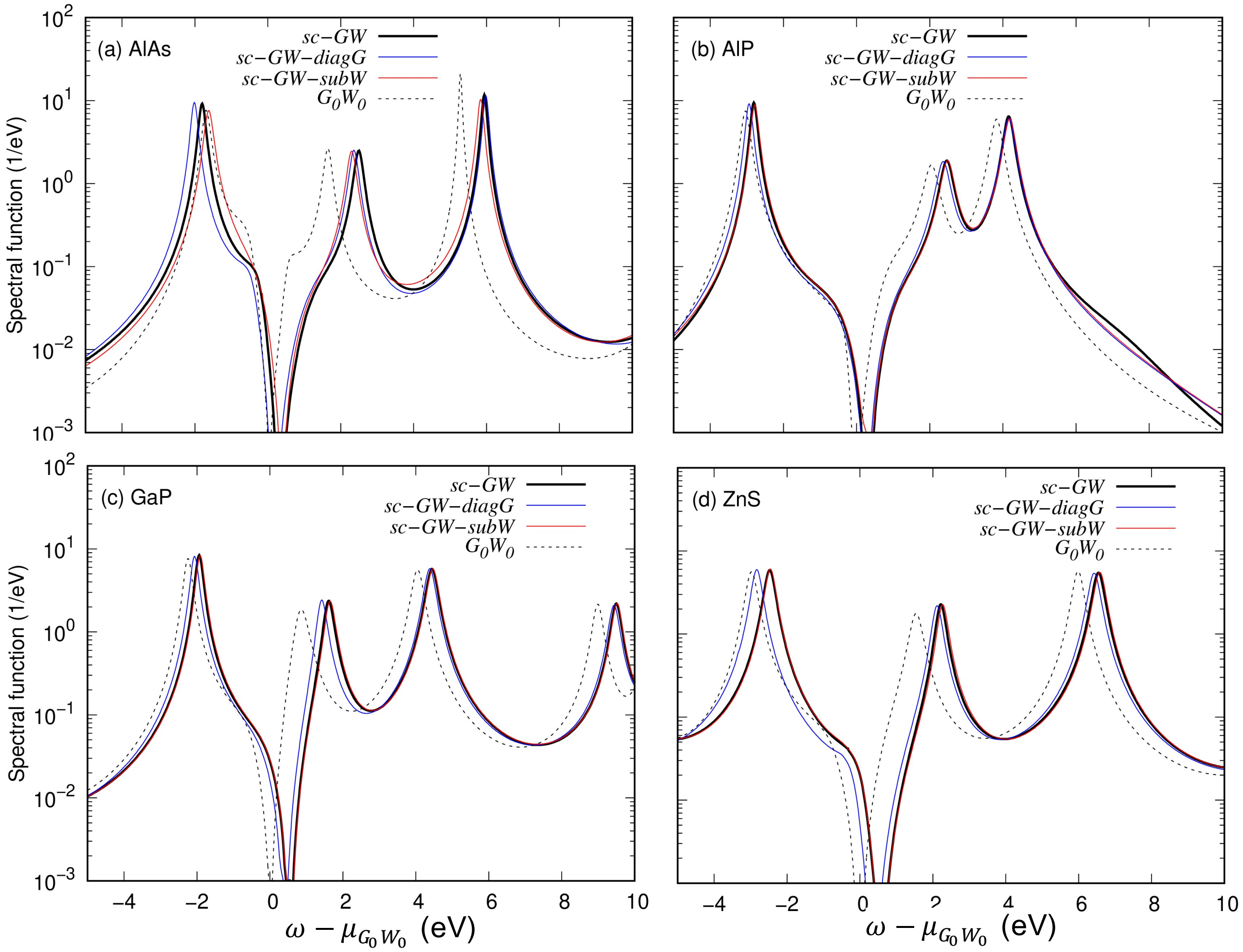

**Figure 3.** A comparison of the spectral functions obtained from $G_0W_0$, *sc-GW*, *sc-GW-diagG,* and *sc-GW-subW* methods for (a) AlAs, (b) AlP, (c) GaP, and, (d) ZnS. The Fermi energy $(\mu_{G_0W_0})$ is at the origin and all other spectrums are relative to that for the comparisons.

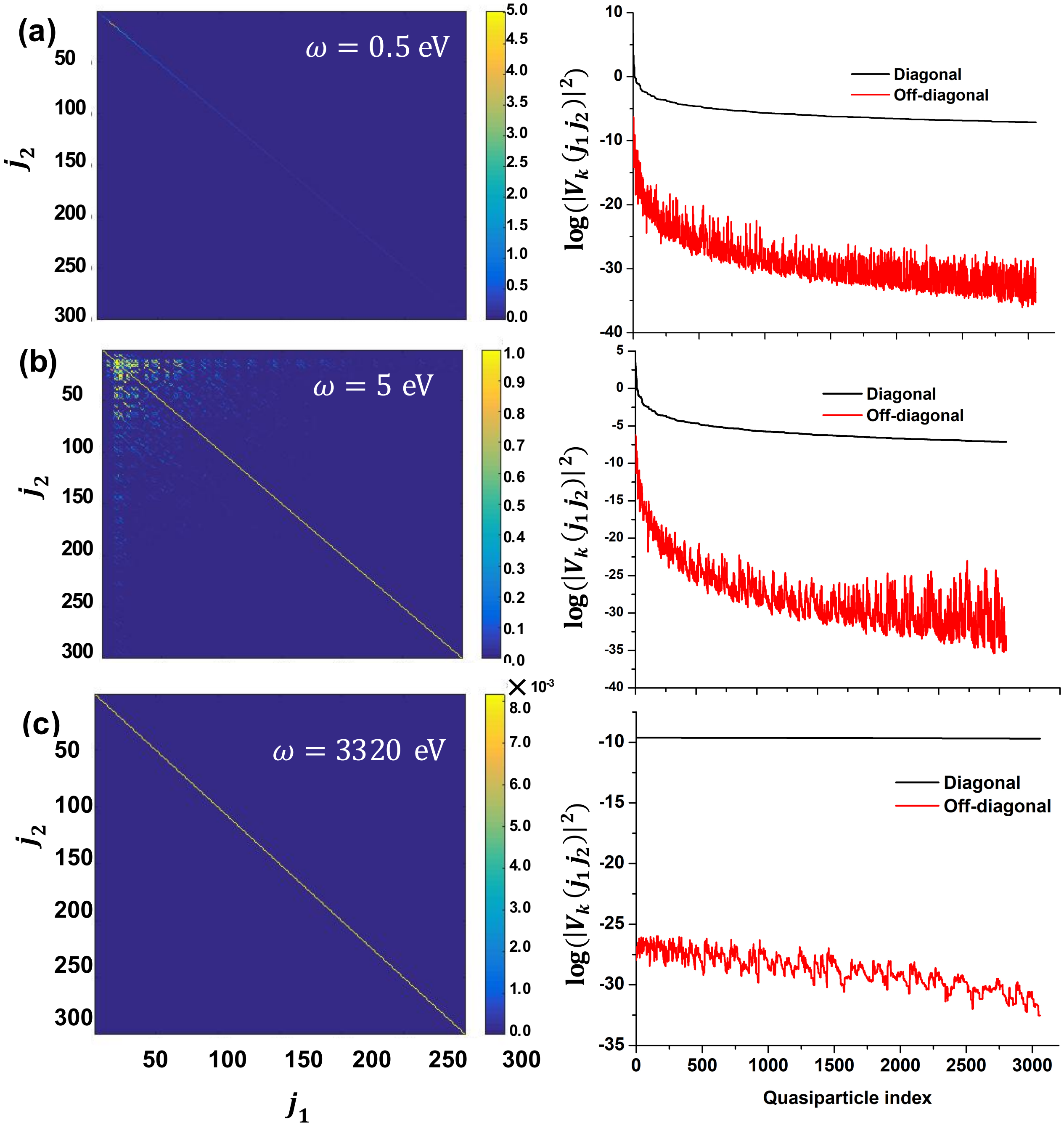

**Figure 4.** Two-dimensional contour plot of the absolute values of a matrix $V_{\boldsymbol{k}}\,(j_1, j_2) = \langle\psi_{j_1\boldsymbol{k}}(\boldsymbol{q_1})|G(\boldsymbol{q_1},\boldsymbol{q_2},\boldsymbol{k}=0,i\omega)|\psi_{j_2\boldsymbol{k}}(\boldsymbol{q_2})\rangle$ at the $\boldsymbol{\Gamma}$ point is shown with the corresponding log of the magnitude of diagonal and off-diagonal terms for AlAs at frequencies (a) $\omega$= 0.5 eV, (b) $\omega$= 5 eV, and (c) $\omega$= 3320 eV. For a given quasiparticle index $j$, in the right-hand side panels, the diagonal terms correspond to the value of $V_k^2(j,j)$, while the off-diagonal term corresponds to $\sum_{j'\neq j} V_k^2\,(j,j')$.

for each (*j*,$\boldsymbol{k}$) state, where, $\epsilon_{j\boldsymbol{k}}(0)$ is the eigenvalue of $H'(\boldsymbol{k},i\omega)$ at zero frequency ($\omega = 0$) and the real part of self-energy is considered to obtain quasiparticle energies. Here, $H'(\boldsymbol{k},i\omega)$ is the Hermitized

Hamiltonian and is given by $[H(\boldsymbol{k}, i\omega) + H^{\dagger}(\boldsymbol{k}, i\omega)]/2$. Furthermore, the spectral function of the periodic solids is related to the trace of the imaginary part of the $G$ matrix as,

$$A(\boldsymbol{k}, \omega) = \frac{1}{\pi}|\mathrm{Tr}\left(\mathrm{Im}\, G(\boldsymbol{k}, \omega)\right)|. \tag{22}$$

With approximation ($|\psi_{j\boldsymbol{k}}(\omega)\rangle \approx |\psi_{j\boldsymbol{k}}(0)\rangle$, this is equivalent to the diagonal approximation as will be discussed below), the above equation can be further transformed to

$$A(\boldsymbol{k}, \omega) = \frac{1}{\pi}\left|\sum_{j} \mathrm{Im}\left[\frac{1}{\omega - \mu - \epsilon_{j\boldsymbol{k}}(0) - [\bar{\Sigma}_{j\boldsymbol{k}}(\omega) - \bar{\Sigma}_{j\boldsymbol{k}}(0)]}\right]\right|. \tag{23}$$

In Figure 2., we plot the analytically extended real and imaginary part of $\bar{\Sigma}(\omega)$ on the real $\omega$ axis for AlAs using our fully *sc-GW*, *sc-GW-diagG,* and *sc-GW-subW* methods. We observe no significant changes in the behavior of $\mathrm{Re}(\bar{\Sigma}(\omega))$ near origin due to the diagonal $G$ and subspace $W$ approximations when compared with the fully *sc-GW* results in the considered systems. The point of intersection of the plot $\mathrm{Re}[\bar{\Sigma}_{j\boldsymbol{k}}(\omega) - \bar{\Sigma}_{j\boldsymbol{k}}(0)]$ with the straight line $\omega + \mu - \epsilon_{j\boldsymbol{k}}(0)$ yields the correction in the quasiparticle energies.

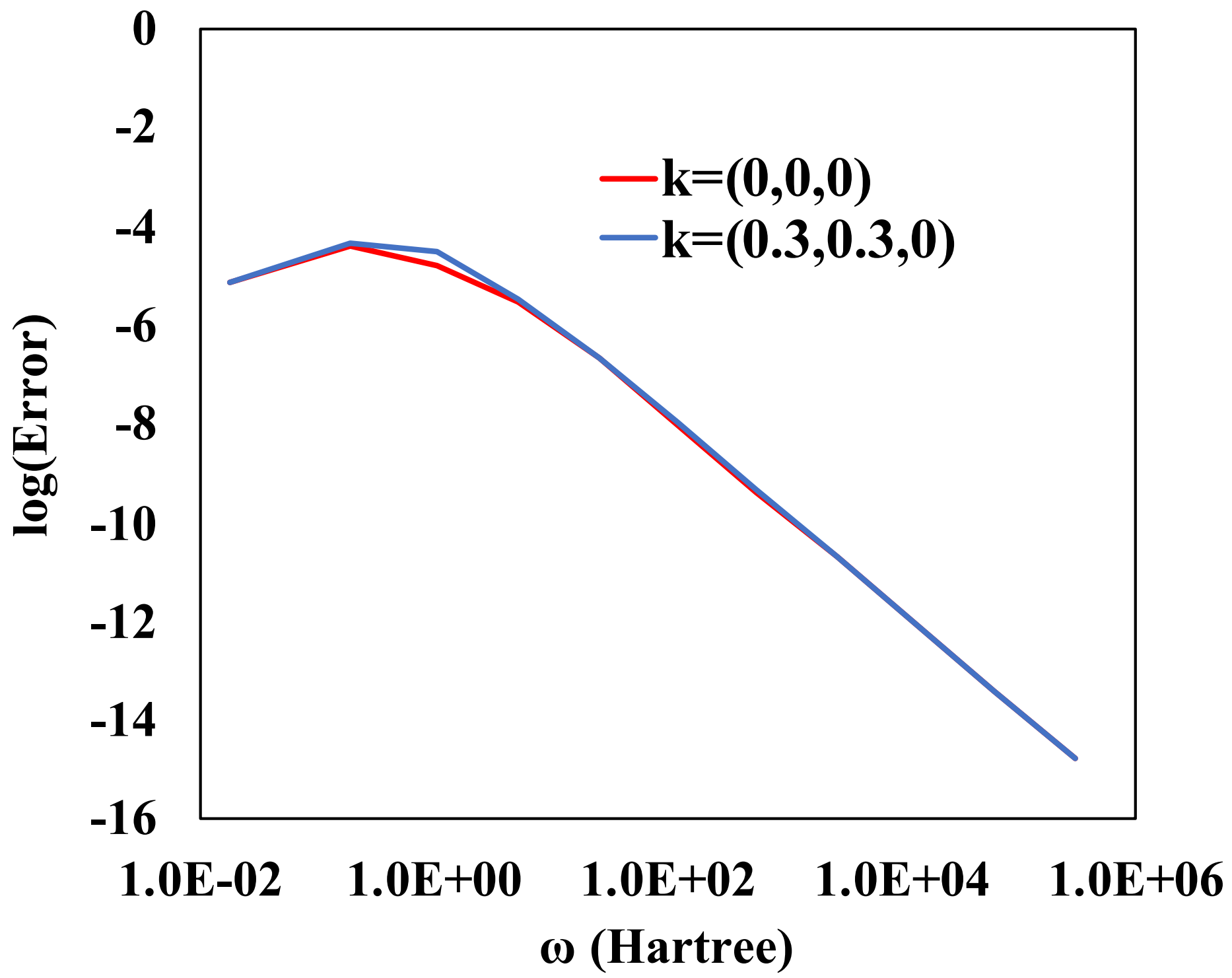

**Figure 5.** Relative error due to ignoring off-diagonal elements of the $G$ matrix as a function of $\omega$ at two different $\boldsymbol{k}$ points, where, Error $= \sum_{\boldsymbol{q_1}\boldsymbol{q_2}} \left| G(\boldsymbol{q_1}, \boldsymbol{q_2}, \boldsymbol{k}, i\omega) - \sum_j \psi_{j,\boldsymbol{k}}(\boldsymbol{q_1})\psi^*_{j,\boldsymbol{k}}(\boldsymbol{q_2}) f_{j,\boldsymbol{k}}(i\omega) \right|^2 / \sum_{\boldsymbol{q_1}\boldsymbol{q_2}} |G(\boldsymbol{q_1}, \boldsymbol{q_2}, \boldsymbol{k}, i\omega)|^2$.

Table I. shows the comparison of calculated quasiparticle bandgaps (at $\boldsymbol{\Gamma}$ point) employing LDA, *G₀W₀*, fully *sc-GW*, *sc-GW-diagG,* and, *sc-GW-subW* methods with other calculations and measurements for bulk AlAs, AlP, GaP, and ZnS systems. The first iteration of our *sc-GW* scheme which is the $G_0W_0$ calculation yields bandgap values for 2.50 eV and 2.53 eV for AlAs and GaP, respectively. These values are approximately 20-30% smaller than the experimental results. In contrast, the $G_0W_0$ bandgap for AlP is overestimated by ~ 8% and for ZnS, this value is in close agreement with the experiment. Thus, compared to the measurements, our $G_0W_0$ bandgap values for the considered system doesn't follow a systematic trend. Further, $G_0W_0$ is highly sensitive to the choice of input DFT wavefunctions and eigenenergies. [14,77] A similar variation in the $G_0W_0$ bandgap values is also observed by Grumet *et al.* within their quasiparticle approximation. Moreover, they used PBE as the exchange-correlation functional in contrast with our LDA for the $G_0W_0$ calculation which is reflected in the bandgap results. [78–81] Nevertheless, overall, $G_0W_0$ bandgaps improve over LDA gap as expected. It is to be noted that, all the earlier works have used bigger approximations in terms of pseudopotentials (e.g., without semi-core), or the truncation of the conduction bands.

Now we turn to self-consistency, which lifts the choice of starting point dependency. [14,17,47,80,82,83] Unlike $G_0W_0$, bandgaps calculated from our fully *sc-GW* method have overestimated the measured bandgaps for all systems. We have also reported similar observations for GaAs, ZnO, and CdS in our previous publication that resulted mainly due to the underestimation of the dielectric constants. [19] The *sc-GW* bandgaps for AlAs, AlP, and ZnS are larger by 17%, 20%, and 12%, respectively. For, GaP the difference in the bandgap calculated from *sc-GW* and experiment is within 5%. The bandgap prediction of the $G_0W_0$ method is better than *sc-GW* for ZnS and AlP. For AlAs and GaP *sc-GW* yield slightly better results, but still with significant overestimations. Furthermore, Grumet *et al.* employed an iterative scheme (*scGW*) similar to our fully *sc-GW* method (but with the diagonal $G$ approximations), resulting in slightly bigger bandgap values for AlP and ZnS. [17] The differences in our results are probably due to, (i) they used an approximately norm-conserving (10% to 20% of norm violation was allowed) projector augmented wave (PAW) pseudopotential [84] in contrast with our fully norm-conserving pseudopotential with semi-core electrons, (ii) to obtain $\boldsymbol{k}$-point grid convergence, $\boldsymbol{k}$-point grids from $(2 \times 2 \times 2)$ to $(6 \times 6 \times 6)$ were tested in $scGW$ scheme to yield the head of the dielectric function, whereas, we have used an interpolation scheme for the band structure from a $(6 \times 6 \times 6)$ to $(126 \times 126 \times 126)$ $\boldsymbol{k}$-point grid for improved $\boldsymbol{k}$-point integration specifically near $\boldsymbol{k}$=0 divergence. One can also notice a difference of 0.82 eV in the $scGW$ bandgap values of ZnS calculated using usual and norm-conserving PAW pseudopotentials, thus, showing the significance of norm-conservation properties. [12,17] It is worth mentioning here that, $GW$ implementations using PAW often carried out without core (and semi-core) electrons in its Greens function representation. [20,85] To investigate the origin of bandgap overestimation due to self-consistency, we obtain converged *sc-GW* macroscopic dielectric constant ($\varepsilon_{GW}$) as shown in Table II. and compare them with the other available calculations and high-frequency measurements. The $\varepsilon_{GW}$ values are significantly smaller compared to the experimental values. [73–75,86] The underestimation of the dielectric constant might be a consequence of the neglect of the vertex term, which can effectively further screen the Coulomb interaction. [19,87]

The spectral function of AlAs, AlP, GaP, and ZnS obtained from *sc-GW* (with and without approximations) and $G_0W_0$ are displayed in Figure 3 for *k*=0, showing peaks at the quasiparticle energies. The peaks lying above and below the Fermi energy (μ) are the quasiparticle bands associated with the valence band maximum (VBM) and conduction band minimum (CBM) of the system, respectively. We see an overall up-shift of the spectral peaks when comparing spectrums obtained from the *sc-GW* and $G_0W_0$ methods. Overall, the increase of the bandgap in *sc-GW* comes mostly due to the upward shift in the CBM. Moreover, in our analysis, we do not observe any satellite peak representing plasmon excitation. Cumulant method post-processing might be necessary to yield such satellite peaks. [88]

One of the main goals of the current work is to test the diagonal $G$ and subspace $W$ approximations. Overall, we found negligible changes in the spectral function due to these approximations as shown in Figure 3. Such insignificant changes are also reflected in the bandgap values obtained from all the three methods (*sc-GW, sc-GW-diagG,* and *sc-GW-subW*). The variation in the bandgap values from the *sc-GW-diagG* method compared to fully *sc-GW* is less than 1.5% except for ZnS (5% bigger bandgap). Moreover, $\varepsilon$ values within the diagonal $G$ approximation are within 3.3% of the fully *sc-GW* method showing a very close agreement. To investigate these two approximations more directly to the matrix $G$ and $W$, we have presented a colored contour plot in Figure 4 showing the absolute values of the matrix element $V_{\boldsymbol{k}}(j_1, j_2) = \langle \psi_{j_1\boldsymbol{k}}(\boldsymbol{q_1}) | G(\boldsymbol{q_1}, \boldsymbol{q_2}, \boldsymbol{k} = 0, i\omega) | \psi_{j_2\boldsymbol{k}}(\boldsymbol{q_2}) \rangle$ for AlAs at $\boldsymbol{\Gamma}$ point for three different frequencies ω= 0.5 eV, ω= 5 eV, and ω= 3320 eV, here, $G$ is the *sc-GW* result. At low frequency (ω= 0.5 eV) the contribution to the matrix element mainly originates from the diagonal terms with the negligible size of off-diagonal elements. Since the Green's function is inversely related to frequency ($G = 1/[i\omega + \mu - H(\boldsymbol{k}) - \Sigma(\boldsymbol{k}, i\omega)]$), we observe decrease in the magnitude of the matrix elements as ω increases. At ω = 3320 eV, the relative amplitude of the diagonal terms is much bigger than the total sum of the off-diagonal terms. Thus, in the diagonal $G$ approximation, the biggest error comes from the intermediate value of ω, e.g., at ω = 5 eV. Figure 4 and Figure 5 show the relative error of the off-diagonal term as a function of ω. As we can see, the biggest error appears at $\omega = 1$ Hartree. But even at that point, the total sum of the off-diagonal term is 10,000 times smaller than the diagonal term, indicating the diagonal $G$ is a very good approximation. This also means the quasiparticle wavefunction $\psi_{j,\boldsymbol{k}}(i\omega)$ is independent of $\omega$. This justifies the use of $\psi_{j,\boldsymbol{k}}(0)$ to calculate the expectation value of $\Sigma(\omega)$ in our spectral function calculation in Eq. (23).

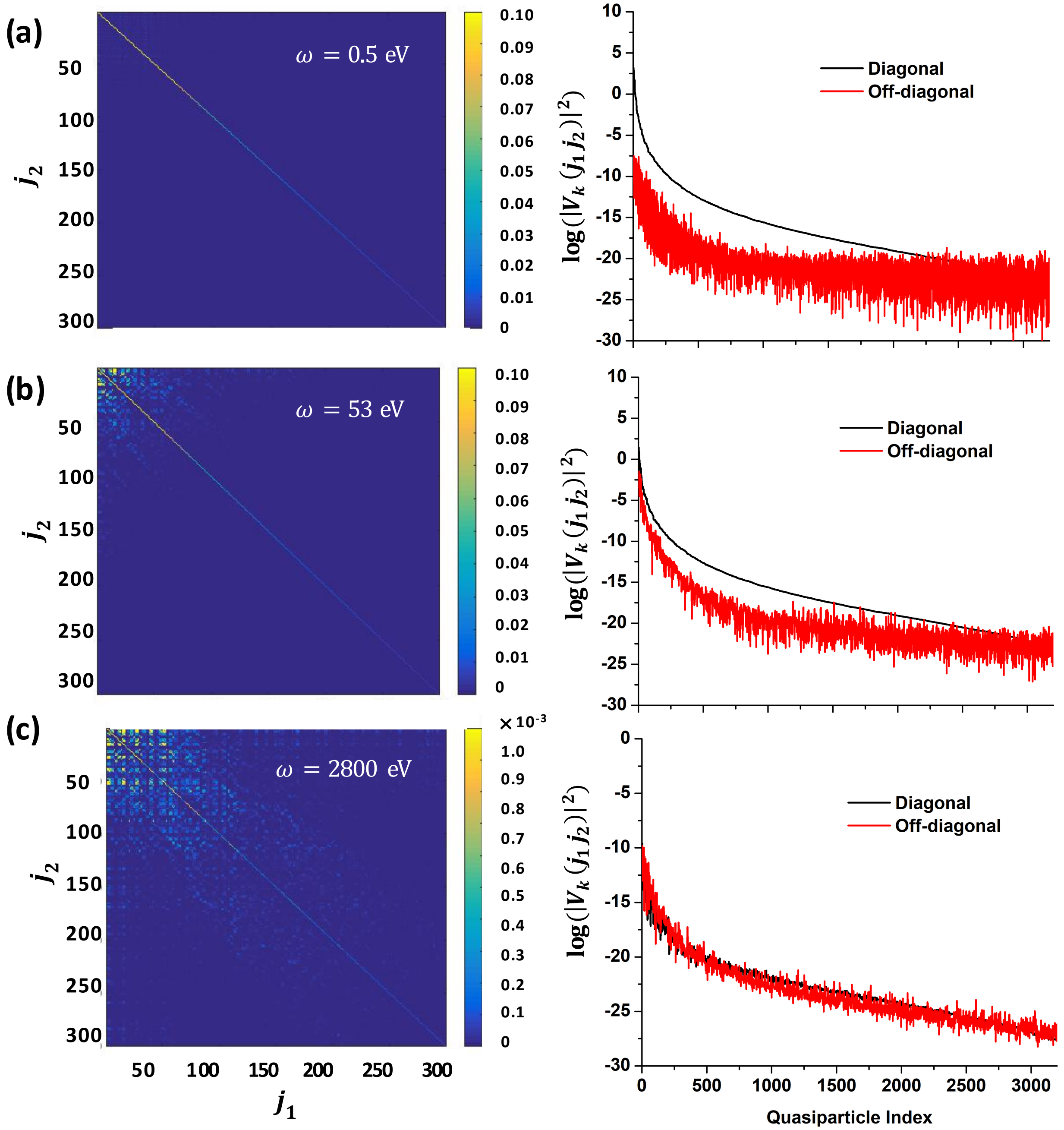

**Figure 6.** Two-dimensional contour plot of the absolute values of a matrix $V(j_1, j_2, \boldsymbol{k} = 0, i\omega) = \langle \theta_{j_1\boldsymbol{k}}(\boldsymbol{q_1}) | W'(\boldsymbol{q_1}, \boldsymbol{q_2}, \boldsymbol{k} = 0, i\omega) | \theta_{j_2\boldsymbol{k}}(\boldsymbol{q_2}) \rangle$ at the $\boldsymbol{\Gamma}$ point is shown with the corresponding log of the magnitudes of diagonal and off-diagonal terms for AlAs at frequencies (a) $\omega$= 0.5 eV, (b) $\omega$= 53 eV, and (c) $\omega$= 2800 eV. The definition of diagonal and off-diagonal terms in the right-hand panels are the same as in Fig.4.

The variations in the bandgap values obtained from the *sc-GW-subW* method compared to fully *sc-GW* is less than 1% with an exception to GaP showing a tiny difference of 1.7%. Further, the $\varepsilon$ values extracted from the *sc-GW-subW* method are within 1.4% of $\varepsilon$ obtained from our fully *sc-GW* approach. In our subspace $W$ approximation, we have taken 30 lowest modes of $W'$. To check the accuracy of this approximation directly on the $W'$ matrix, we perform an analysis similar to above by plotting a colored contour map of the absolute values of the matrix element $V(j_1, j_2, \boldsymbol{k} = 0, i\omega) = \langle \theta_{j_1\boldsymbol{k}}(\boldsymbol{q_1}) | W'(\boldsymbol{q_1}, \boldsymbol{q_2}, \boldsymbol{k} = 0, i\omega) | \theta_{j_2\boldsymbol{k}}(\boldsymbol{q_2}) \rangle$ at $\boldsymbol{\Gamma}$ point under the basis set of the eigenmodes of $W'(\omega = 0)$ for $\omega = 0.5$ eV, $\omega =$ 53 eV, and $\omega = 2800$ eV (Figure 6). At $\omega = 0.5$ eV, the diagonal elements dominate, but the off-diagonal terms are not zero, again due to the non-Hermitian property of $W'$ at non-zero $\omega$ values. The dominance of the diagonal term disappears as the $\omega$ increases. Thus, a simple diagonal approximation for $W'$ will not work. On the other hand, the values of elements drop exponentially as a function of the index of the eigenmodes. Thus, one can truncate the index, and make a low-rank approximation of $W'$. At a very large $\omega$, the whole value of $W'$ is so small, one can even ignore the $W'$ altogether. Note the rank of $W'$ is about 8000, thus reducing it to 30 is a dramatic reduction in terms of both memory and computational cost.

## 4. Conclusions

To summarize, we presented quasiparticle energies and spectral function for bulk AlAs, AlP, GaP, and ZnS from a fully self-consistent $GW$ calculation (*sc-GW*) based on the Dyson equation using plane-wave basis set. Norm-conserving semi-core pseudopotentials with LDA is employed for the starting DFT computations. A full $G$ matrix is constructed without any conduction band truncation. We observed that the initial non-self-consistent $G_0W_0$ results increase the LDA bandgaps and show positive as well as negative differences when compared with the experimental results. On the other hand, self-consistence *sc-GW* always leads to overestimated bandgaps mainly due to underestimation in the macroscopic dielectric constant ($\varepsilon$). From the spectral function, we observe that *sc-GW* bandgap widens mostly by shifting the $G_0W_0$ conduction band furthers up in energy during self-consistent iterations.

One of the main goals in the current work is to test the diagonal $G$ approximation and low-rank $W$ approximations widely used in the literature. We implement diagonal $G$ (*sc-GW-diagG*) and subspace $W$ (*sc-GW-subW*) approximations within the frameworks of our fully *sc-GW* scheme. The tests are carried out in two different ways. First, using such approximations, the self-consistent iterations are carried out, and the final quasiparticle bandgap, as well as the spectral functions, are compared with the *sc-GW* result. Second, the approximated diagonal $G$ and low-rank $W'$ are directly compared to the full matrix $G$ and $W'$ of the *sc-GW* calculations. We found that, the *sc-GW-diagG* and *sc-GW-subW* methods predict the *sc-GW* quasiparticle bandgaps within 1.7% except for ZnS whose bandgap value with diagonal-*G* approximation differ by approximately 5% with $sc\text{-}GW$. Furthermore, the $\varepsilon$ value obtained from $sc\text{-}GW\text{-}diagG$ and $sc\text{-}GW\text{-}subW$ methods differ from the fully $sc\text{-}GW$ calculation by less than 3.3% and 1.4%, respectively. Using the direct matrix comparison, we found that the sum of all the off-diagonal elements in the full matrix $G$ is less than $10^{-4}$ of the diagonal value of $G$ for all values of $\omega$. The largest error comes from the intermediate value of $\omega$. As for the $W'$, its matrix elements decrease exponentially as a function of the eigenmode index. This allows us to truncate the matrix, from a rank 8,000 to rank 30 matrix, without introducing big errors. Our test validates the widely used diagonal approximation for $G$ and low-rank approximation of $W'$. Such approximations can serve as the starting point to incorporate vertex corrections based on the Feynman diagram.

**Acknowledgment**

This work was supported by the Center for Computational Study of Excited-State Phenomena in Energy Materials at the Lawrence Berkeley National Laboratory, which is funded by the U.S. Department of Energy, Office of Science, Basic Energy Sciences, Materials Sciences and Engineering Division under Contract No. DE-AC02-05CH11231, as part of the Computational Materials Sciences Program. This work used the resources of the National Energy Research Scientific Computing Center (NERSC) and Oak Ridge Leadership Computing Facility (OLCF) with the computational time allocated by the INCITE program.